\documentstyle{article}
\newcommand{\be}{\begin{equation}}
\newcommand{\ee}{\end{equation}}
\newcommand{\bea}{\begin{eqnarray}}
\newcommand{\eea}{\end{eqnarray}}
\newcommand{\ba}{\begin{array}}
\newcommand{\ea}{\end{array}}

\newcommand{\hs}[1]{\hspace{#1 mm}}
\def\bbox{{\,\lower0.9pt\vbox{\hrule \hbox{\vrule height 0.2 cm
\hskip 0.2 cm \vrule height 0.2 cm}\hrule}\,}}
\newcommand{\dsl}{\pa \kern-0.5em /}

\newcommand{\nn}{\nonumber \\}

\newcommand{\EQ}{\begin{equation}}
\newcommand{\EN}{\end{equation}}

\newcommand{\e}{\epsilon}
\def\bbox{{\,\lower0.9pt\vbox{\hrule \hbox{\vrule height 0.2 cm
\hskip 0.2 cm \vrule height 0.2 cm}\hrule}\,}}

\newcommand{\pa}{\partial}

\newcommand{\p}[1]{(\ref{#1})}

\def\today{\ifcase\month\or
  January\or February\or March\or April\or May\or June\or
  July\or August\or September\or October\or November\or December\fi
 \space\number\day, \number\year}

\input amssym.def
\input amssym.tex
\font\mybb=msbm10 at 10pt
\def\bb#1{\hbox{\mybb#1}}

\def\bR {\bb{R}}
\def\bE {\bb{E}}

\begin{document}

%%%%%%%%%%%%%%%% title page %%%%%%%%%%%%%%%%%%%%%%%%%%%%%%%%%%%%

\begin{titlepage}
\vfill
\begin{flushright}
UG-00-11\\
OU-HET 362 \\
DAMTP-2000-103\\
hep-th/0009147\\
\end{flushright}

\vfill
\begin{center}
\baselineskip=16pt
{\Large\bf M-BRANE INTERPOLATIONS AND (2,0) RENORMALIZATION GROUP FLOW}
\vskip 0.3cm
{\large {\sl }}
\vskip 10.mm
{\bf ~Eric Bergshoeff$^{*,1}$, ~Rong-Gen Cai$^{\sharp,2}$,\\
Nobuyoshi Ohta$^{\sharp,3}$ and  ~Paul K. Townsend$^{\dagger,4}$ } \\
%\\[2mm]
\vskip 1cm
%\vfill
{\small
$^*$
 Institute for Theoretical Physics, University of Groningen, \\
Nijenborgh 4, 9747 AG Groningen, The Netherlands\\
}
\vspace{6pt}
{\small
 $^\sharp$
Department of Physics, Osaka University, \\
Toyonaka, Osaka 560-0043, Japan\\
}
\vspace{6pt}
{\small
 $^\dagger$
DAMTP, University of Cambridge, \\
Centre for Mathematical Sciences,
Wilberforce Road, \\
Cambridge CB3 0WA, UK\\
}
\end{center}
\vfill
\par

\begin{center}
{\bf ABSTRACT}
\end{center}
\begin{quote}

We obtain the M5-M2-MW bound state solutions of 11-dimensional supergravity
corresponding to the 1/2 supersymmetric vacua of the M5-brane equations with
constant background fields. In the `near-horizon' case the solution
interpolates between the $adS_7\times S^4$ Kaluza-Klein vacuum and
D=11 Minkowski spacetime via a Domain Wall spacetime. We discuss
implications for renormalization group flow of (2,0) D=6 field
theories.

\vfill
 \hrule width 5.cm
\vskip 2.mm
{\small
\noindent $^1$ E-mail: bergshoe@phys.rug.nl \\
\noindent $^2$ E-mail: cai@het.phys.sci.osaka-u.ac.jp\\
\noindent $^3$ E-mail: ohta@phys.sci.osaka-u.ac.jp\\
\noindent $^4$ E-mail: p.k.townsend@damtp.cam.ac.uk \\
}
\end{quote}
\end{titlepage}

%%%%%%%%%%%%%%%%%%%%%%%%%%%%%%%%%%%%%%%%
\setcounter{equation}{0}
\section{Introduction}

Following many studies of D-branes in constant $B$-field backgrounds, a number
of papers have considered the analogous problem of the M5-brane in a
constant background 3-form gauge field $C$ (e.g. \cite{M}).
The background field $C$ appears, via its pullback, in the M5-brane
action through the worldvolume 3-form field strength $H= dA - C$.
An M5-brane in a {\sl constant} $C$
background is therefore equivalent to an M5-brane in the M-theory vacuum but
with constant worldvolume 3-form $H$. There is a class of such
constant M5-brane
configurations that preserve all 16 supersymmetries of the M5-brane vacuum
\cite{ST}. This class is characterized by the M5-brane charge $y$, the
M-Wave (MW) charge p, which is a momentum
in one direction in the M5-brane, and the skew eigenvalues $\xi_1,\xi_2$ of
the spatial components of $H$ in the 4-directions orthogonal to $p$; these
skew eigenvalues are M2-brane charges, as one discovers by a computation of the
supersymmetry algebra of the M5-brane Noether charges \cite{ST}.
This calculation leads to the conclusion, for a particular choice of
worldvolume coordinates and assuming that the tension is normalized to
unity, that all 16 worldvolume supersymmetries will be
preserved if and only if the equation
\bea
\Big( \xi_1 \Gamma_{012} + \xi_2 \Gamma_{034} + p \Gamma_{05}
+ y \Gamma_{012345} \Big) \e = \e
\eea
admits 16 linearly-independent non-zero solutions for the constant real D=11
spinor $\epsilon$. This is the case if and only if the charges
$(y,p,\xi_1,\xi_2)$ satisfy
\bea\label{relations}
\xi_1^2 + \xi_2^2 +p^2 + y^2 = 1, \nn
\xi_1 \xi_2 - py=0.
\eea
The solutions of these constraints can be parametrized by two angles
$(\theta_1,\theta_2)$ as follows:
\bea
\xi_1 &=& s_1 c_2,\quad
\xi_2 = c_1 s_2, \nn
p &=& s_1 s_2,\quad
y = c_1 c_2.
\label{sol}
\eea
where
\be\label{angles}
s_i =\sin\theta_i \,, \qquad c_i=\cos\theta_i\,, \qquad (i=1,2).
\ee
Thus, there is a two-parameter family of M5-brane vacua. They are vacua in the
sense that they preserve all 16 supersymmetries, although they differ in
energy because they minimise the energy subject to different boundary
conditions (namely that the fields approach their prescribed constant values at
infinity).

The above result was found from the action of a single M5-brane.
The action for multiple M5-branes is not known but the low energy
dynamics is presumed to be governed by a (2,0)-supersymmetric D=6
superconfomal field theory (see e.g. \cite{seiberg}). It is thus natural
to suppose that this theory is a particular, superconformal,
member of a 2-parameter family of (2,0) field theories,
parameterized as above. The superconformal (2,0) theory is believed to be
equivalent, via the adS/CFT correspondence, to M-theory on
$adS_7\times S^4$, which is the `near-horizon' limit of the D=11 supergravity
M5-brane. The other, non-conformal, (2,0) field theories are then
presumably equivalent to M-theory in a background that is the
near-horizon limit of a bound state solution of an M5-brane with
M2-branes and an M-Wave, the charges of these constituents
being related in the way described above. We thus expect there to
exist a family of 1/2 supersymmetric M5-M2-MW bound state solutions
of D=11 supergravity parameterized by the two angles
$(\theta_1,\theta_2)$. One purpose of this paper is to present this
family of solutions, which we obtain using the methods of
\cite{BMM,CP}. The solutions are essentially the lift to D=11 of
D-brane bound state solutions found in \cite{BMM,HO}. Each depends on a
single harmonic function $H$ on $\bE^5$, the space transverse to the
M5-brane's worldvolume in eleven dimensions. The full stationary
D=11 solution has not previously been given, although the static
M5-M2-brane bound state solutions were found in \cite{ILPT} and the
M2-MW case is the boosted M2-brane of \cite{RT}.

We shall be interested here in the `near-horizon' case obtained by choosing
\be\label{Hchoice}
H= 1/r^3
\ee
where $r$ is the radial distance from the brane in the transverse $\bE^5$
space. For this choice, the pure M5-brane solution becomes
the $adS_7\times S^4$ Kaluza-Klein vacuum \cite{GT}, with isometry group
\be
SO(2,6)\times SO(5)
\ee
as expected by the equivalence of M-theory in this background to the conformal
(2,0) theory. In the generic M5-M2-MW case the isometry group is
\be\label{isom}
\bR^2 \times ISO(2)\times ISO(2) \times SO(5)
\ee
but this is typically enhanced in either of the limits $r\rightarrow 0$ or
$r\rightarrow \infty$. In particular, it is always enhanced to $SO(2,6)\times
SO(5)$ in the $r\rightarrow 0$ limit, provided that the M5 charge is non-zero;
this generalizes the observation of \cite{ILPT} that the M5-brane
dominates the M5-M2 solution in this limit.
For the pure M5-brane the $r\rightarrow\infty$ limit yields the same
as the $r\rightarrow 0$ limit (for $H=1/r^3$). This was to be expected
from the conjectured equivalence of M-theory in the
near-horizon M5-brane background with the  {\sl superconformal}
(2,0) theory. In all other cases the $r\rightarrow \infty$ limit
yields an asymptotic spacetime that is {\sl not} $adS_7\times S^4$.
We interpret this to mean that the non-conformal (2,0) theories flow to
the conformal (2,0) theory in the IR limit.

Of particular interest is a limit in the two-parameter space corresponding to a
critical electric component of $H$, as this has been argued to lead to an Open
Membrane (OM) theory \cite{gop,BBS}. Because of the non-linear self-duality
condition obeyed by $H$, this limit corresponds to one in which some
magnetic components of $H$ go to infinity. This implies that the
M5-brane tension must also go to infinity, and if one rescales to keep the
tension at unity then the limit is one
in which either $\xi_1$ or $\xi_2$ becomes large relative to $y$. The only way
the relations (\ref{relations}) can be satisfied in this limit is if either
$\xi_1 \rightarrow 1$ or $\xi_2 \rightarrow 1$ (but not both). The
supergravity dual in this limit was studied in \cite{BS} for the
special case in which $\xi_2=p=0$, using a form of the static M5-M2 bound state
solution of D=11 supergravity found in \cite{CGHN}. An asymptotic `smeared
membrane' spacetime was found and argued to be the background associated with
the supergravity dual to OM-theory.

Part of the motivation for the work reported here was to get a better
understanding of the renormalization group (RG) flow to the conformal (2,0)
theory by considering the general M5-M2-MW solution and its interpolation
properties. For the special case of the M5-M2 bound state we find an
interpolation between the $adS_7\times S^4$ vacuum (for
$r\rightarrow 0$) and (for $r\rightarrow\infty$) the near-horizon limit of the
M2-brane as a solution of the maximal D=8 supergravity \cite{BST}, for which
the `dual-frame' 8-metric is $adS_4\times S^4$. This solution was first
obtained as a D=4 domain wall (DW) solution
of ($T^3\times S^4$)-compactified D=11
supergravity \cite{Cowdall}, so we shall refer to it as the DW solution.
The generic M5-M2-MW solution, however, has quite different interpolation
properties. When the MW charge is non-zero the metric is asymptotic, as
$r\rightarrow \infty$, to a flat D=11 vacuum spacetime. This is achieved
via an intermediate DW spacetime.

We begin our presentation of these results with the construction of the general
1/2 supersymmetric M5-M2-MW solution of D=11 supergravity, which we obtain by a
series of solution-generating manipulations from the D2-brane solution of IIA
D=10 supergravity. We then specialize to the `near-horizon' choice
(\ref{Hchoice}) of harmonic function, and consider the $r\rightarrow 0$ and
$r\rightarrow \infty$ limits. We conclude with a summary of the RG
interpretation and a discussion of some related issues.

\section{Construction}

We start from the D2-brane solution of D=10 IIA supergravity
\bea
ds_A^2 &=& H^{-1/2} ( -dt^2 + dx_1^2 + dx_3^2) + H^{1/2} ( dx_2^2 + dx_4^2
 + dr^2 +r^2 d\Omega_4^2),\nn
&& \phi = \frac{1}{4} \log H,\quad
C = \frac{1-H}{H} dt \wedge dx_1 \wedge dx_3,
\label{1}
\eea
where $H$ is a harmonic function on the transverse space, which we shall take
to be independent of $x_2$ and $x_4$; in other words, we have a D2-brane
`smeared' in the $x_2$ and $x_4$ directions, which we assume are compact.
We now define new rotated coordinates $(\tilde x_1,\tilde x_2,\tilde x_3,\tilde
x_4)$ by
\bea
\left( \begin{array}{l}
x_1 \\
x_2
\end{array} \right)
=
\left( \begin{array}{cc}
c_1 & -s_1 \\
s_1 & c_1
\end{array} \right)
\left( \begin{array}{l}
{\tilde x}_1 \\
{\tilde x}_2
\end{array} \right), \quad
\left( \begin{array}{l}
x_3 \\
x_4
\end{array} \right)
=
\left( \begin{array}{cc}
c_2 & -s_2 \\
s_2 & c_2
\end{array} \right)
\left( \begin{array}{l}
{\tilde x}_3 \\
{\tilde x}_4
\end{array} \right),
\eea
where $s_i$ and $c_i$ are the sines and cosines of (\ref{angles}).
In the new coordinates the IIA solution \p{1} is
\bea
ds_A^2 &=& -H^{-1/2} dt^2 + H^{-1/2}(c^2_1 +H s^2_1 ) dx_1^2
 + H^{-1/2}(s^2_1 +H c^2_1 ) dx_2^2 \nn
&& +2H^{-1/2}(H-1)c_1s_1 dx_1dx_2
 + H^{-1/2}(c^2_2 +H s^2_2 ) dx_3^2 \nn
&& + H^{-1/2}(s^2_2 +H c^2_2) dx_4^2 +2H^{-1/2}(H-1)
c_2s_2 dx_3 dx_4
\nn
&& + H^{1/2}[dr^2 +r^2 d\Omega_4^2], \nn
&& \hs{-15} \phi = \frac{1}{4} \log H,\quad
C = \frac{1-H}{H} dt \wedge (dx_1c_1-dx_2s_1)
\wedge (dx_3c_2-dx_4s_2),
\label{2}
\eea
where we have now dropped the tildes.

Performing a T-duality in the $x_2$-direction, we obtain the IIB supergravity
solution
\bea
ds_B^2 &=& -H^{-1/2} dt^2 + \frac{H^{1/2}}{E_1} (dx_1^2 + dx_2^2 )
 + H^{-1/2}(c^2_2 +H s^2_2) dx_3^2 \nn
&& + H^{-1/2}(s^2_2 +H c^2_2) dx_4^2
 +2H^{-1/2}(H-1)c_2s_2 dx_3 dx_4 \nn
&& + H^{1/2}[dr^2 +r^2 d\Omega_4^2], \nn
\varphi &=& {1\over 2}\log \frac{H}{E_1}, \nn
D &=& \frac{1-H}{E_1} c_1 dt \wedge dx_1 \wedge dx_2 \wedge
(dx_3 c_2-dx_4 s_2), \nn
B^{(1)} &=& \frac{H-1}{E_1} c_1s_1 dx_1\wedge dx_2, \nn
B^{(2)} &=& \frac{1-H}{H} dt\wedge (dx_3 c_2 -dx_4 s_2)s_1\, ,
\label{3}
\eea
and a further T-duality in the $x_4$-direction converts this to the IIA
supergravity solution
\bea
ds_A^2 &=& H^{1/2} \left[ -H^{-1} dt^2 + \frac{1}{E_1} (dx_1^2 + dx_2^2 )
 + \frac{1}{E_2} (dx_3^2 + dx_4^2 ) + dr^2 +r^2 d\Omega_4^2 \right], \nn
\phi &=& \log H^{3/4}E_1^{-1/2}E_2^{-1/2}, \nn
A &=& \frac{1-H}{H} s_1s_2 dt, \nn
B &=& \frac{H-1}{E_1} c_1s_1 dx_1\wedge dx_2
 + \frac{H-1}{E_2} c_2s_2 dx_3\wedge dx_4, \nn
dC &=& d\Big(\frac{1-H}{E_1}\Big) c_1s_2 \wedge dt\wedge dx_1 \wedge dx_2
 + d\Big(\frac{1-H}{E_2}\Big) c_2s_1 \wedge dt\wedge dx_3 \wedge dx_4
 - c_1c_2 \star dH\, , \nn
\label{4}
\eea
where
\be
E_1 = s^2_1 + Hc^2_1\, ,\qquad E_2 = s^2_2 + Hc^2_2\, .
\ee
This is the desired D0-D2-D2-D4 brane solution~\cite{BMM,HO}.

Uplifting to 11 dimensions, we get the following new 1/2 supersymmetric
solution of D=11 supergravity :
\bea
ds_{11}^2 &=& (E_1 E_2)^{1/3} \Big[ -H^{-1}\left[1
 -(1-H)^2 E_1^{-1} E_2^{-1} s^2_1 s^2_2 \right] dt^2 \nn
&& + 2 E_1^{-1} E_2^{-1} (1-H)  s_1 s_2 dt dx_\natural
 + H E_1^{-1} E_2^{-1} dx_\natural^2 \nn
&& + E_1^{-1} (dx_1^2 + dx_2^2 ) + E_2^{-1} (dx_3^2 + dx_4^2 )
 + dr^2 +r^2 d\Omega_4^2 \Big], \nn
dC &=& d\Big(\frac{1-H}{E_1}\Big) c_1 s_2 \wedge
 dt \wedge dx_1 \wedge dx_2
 + d\Big(\frac{1-H}{E_2}\Big) c_2 s_1 \wedge dt \wedge dx_3
 \wedge dx_4 \nn
&& \hs{-10}+ d\Big(\frac{H-1}{E_1}\Big) c_1 s_1  \wedge dx_1 \wedge
 dx_2 \wedge dx_\natural
 + d\Big(\frac{H-1}{E_2}\Big) c_2 s_2 \wedge dx_3 \wedge dx_4
 \wedge dx_\natural \nn
&& - c_1c_2 \star d H\, ,
\eea
When $s_1s_2=0$ this reduces to the M5-M2 brane solution of \cite{ILPT}; the
subcase with $s_1=s_2=0$ is the pure M5-brane. When $c_1c_2=0$ it is
the boosted membrane solution of \cite{RT}; the subcase with
$c_1=c_2=0$ is the pure M-wave solution. The general case is a bound
state solution of an M5-brane with an M-Wave and two orthogonal M2-branes.
It has the isometry group (\ref{isom}); the $\bR^2$ factor is
generated by the Killing vector fields $\partial/\partial t$ and
$\partial/\partial x_\natural$. Although $\partial/\partial t$ is not
timelike for all $r$, the Killing vector field
\be
k = {\partial\over \partial t} + s_1s_2 {\partial\over \partial x_\natural}\, ,
\ee
is. For this reason it is convenient to define a new space
coordinate $\tilde x$ by
\be
\tilde x = x_\natural - s_1s_2 t\, .
\ee
In the new coordinates $k=\partial/\partial t$. The metric is
\bea\label{newmet}
ds^2 &=& (E_1 E_2)^{-{2\over3}}\bigg\{ -\left(Hc_1^2 c_2^2 + c_1^2 s_2^2 +
c_2^2 s_1^2\right) dt^2 + 2s_1s_2 dtd\tilde x + Hd\tilde x^2 \nn
&&+ \,
E_2 \left(dx_1^2 + dx_2^2\right) + E_1\left(dx_3^2 + dx_4^2\right)
+ E_1 E_2 \left(dr^2 + r^2d\Omega_4^2\right)\bigg\}\, ,
\eea
and the 4-form field strength is
\bea
&&\hs{-5} F = -dH\wedge\left[c_1 E_1^{-2}\left(s_2 c_1^2 dt - s_1
 d\tilde x\right) dx_1 \wedge dx_2
+ c_2 E_2^{-2}\left(s_1 c_2^2 dt - s_2 d\tilde x\right)dx_3 \wedge dx_4
 \right]\nn
&& \hs{5} - \, c_1c_2 \star dH\, .
\eea
This will be the starting point for the analysis to follow.

\section{Interpolations}

A simple choice of the harmonic function $H$ in the M5-M2-MW solution is
$H = a+ 1/ r^3$ for non-negative constant $a$. When $a>0$ the solution is
asymptotically flat. Here we shall be interested in the behaviour of the
`near-horizon' solution with $a=0$; that is, with $H=1/r^3$.

We begin by examining the behaviour as $r\rightarrow 0$. Provided
$c_1c_2$ is non-zero (i.e. non-zero M5-charge) we find the asymptotic solution
\bea
ds^2 &=& (c_1c_2)^{2\over3}\bigg\{ r\left[-dt^2 + (c_1c_2)^{-2}d\tilde x^2 +
c_2^{-2}\left( dx_1^2 + dx_2^2\right)+  c_1^{-2}\left( dx_3^2 + dx_4^2\right)
\right]\nn
&&+\, r^{-2}dr^2 + d\Omega_4^2\bigg\} \, ,\nn
F&=& 3c_1c_2 r^{-4}\star dr\, ,
\eea
which is the $adS_7\times S^4$ Kaluza-Klein vacuum. This is exactly the same
as the near-horizon limit of the pure M5-brane solution \cite{GT}. Thus the
M5-brane `dominates' as $r\rightarrow0$. This result was found previously for
the static M5-M2 solution in \cite{ILPT}; we now see that it is true for the
general stationary M5-M2-MW solution.

In the special case that the M5-brane charge vanishes ($c_1c_2=0$) it is the
M2-brane which dominates in the $r\rightarrow 0$ limit. To see this we set
$c_2=0$, in which case
\be
E_2=1\, ,\qquad E_1 = E \equiv s^2 + Hc^2\, ,
\ee
and the solution (\ref{newmet}) reduces to
\bea
ds^2 &=& E^{-{2\over3}}\left\{ -c^2 dt^2 + 2sdtd\tilde x +
Hd\tilde x^2 + dx_1^2 + dx_2^2 + E\left[dx_3^2 + dx_4^2 + dr^2 +
r^2 d\Omega_4^2\right]\right\}\, ,\nn
F &=& -c E^{-2} dH \wedge \{ c^2 dt - s d{\tilde x} \} \wedge dx_1
 \wedge dx_2\, .
\eea
In the limit $r\rightarrow0$ this becomes
\bea
ds^2 &\sim& c^{-4/3}r\left\{ r\left[ -c^2d\tau^2 + dx_1^2 + dx_2^2\right]
+ c^2 r^{-2} dr^2  + c^2 d\Omega_4^2\right\}\nn
&&+\, c^{-4/3} r^{-1} \left[d\tilde x^2 + c^2 \left(dx_3^2 + dx_4^2\right)
\right]\, \nn
F &\sim& 3 c^{-1} r^2 dr \wedge d\tau \wedge dx_1 \wedge dx_2\, ,
\eea
where
\be
\tau = t- {s\over c^2} \tilde x\, .
\ee
This is just the near-horizon limit of the M2-brane, as a solution of the
$T^3$-compactified D=11 supergravity; the 8-metric in the curly
parenthesis is the `dual-frame' $adS_4 \times S^4$ 8-metric \cite{BST}.
In summary, there is a `dominance' hierarchy among the M5, M2 and MW
components in
the $r\rightarrow 0$ limit with the M5 dominating the M2 and MW and the M2
dominating the MW. It follows that the M5-brane dominates the M-Wave, which
means that the singularity of the pure M-Wave solution is removed when it is
part of the generic M5-M2-MW solution.

We now turn to the $r\rightarrow \infty$ limit. We shall
begin with the static M5-M2 case by setting $s_2=0$. In this case
\be
E_1= E\equiv s^2 + Hc^2\, ,\qquad E_2 =H\, ,
\ee
and, assuming that $s$ is non-zero, the asymptotic solution is
\bea
ds^2 &=& s^{2\over3} r \left\{ r\left[-dt^2 + dx_3^2 + dx_4^2\right] + r^{-2}
dr^2 + d\Omega_4^2 + s^{-2}r^{-2}\left[d\tilde x^2 + dx_1^2
+ dx_2^2\right]\right\}\nn
F &\sim& 3 s r^2 dr \wedge dt \wedge dx_3 \wedge dx_4\, .
\eea
Remarkably, this is the same (after some trivial rescaling of coordinates) as
the solution found above in the $r\rightarrow 0$ limit of the M2/MW solution.
This result depends crucially on $s\ne0$ (and, of course, on $a=0$).
When $s=0$, we have the
pure M5-brane solution for which the $r\rightarrow \infty$ and
$r\rightarrow 0$ limits are identical. The `mixed' M5-M2 case is thus
quite different. We shall discuss the significance of this below, but
here we may remark that it implies a `dominance' of the M2 over the M5
in the $r\rightarrow\infty$ limit.

We now turn to the $r\rightarrow\infty$ limit of the
generic stationary solution with non-zero $s_1s_2$. Defining
\be
\tau = t - (s_1s_2A)^{-1}\tilde x\, ,\qquad
A= {c_1^2\over s_1^2} + {c_2^2 \over s_2^2} \, ,
\ee
we find in this case that
\bea
ds^2 &\sim&  (s_1s_2)^{2\over3} \left\{ -Ad\tau^2 + {1\over
As_1^2s_2^2}d\tilde x^2 + {1\over s_1^2} \left(dx_1^2 + dx_2^2\right)
+ {1\over s_2^2}\left(dx_3^2 + dx_4^2\right) + dr^2 + r^2 d\Omega_4^2
\right\} \nn
F &\sim& 0\, .
\eea
This is a flat vacuum solution of D=11 supergravity, which is
remarkable given that we are discussing the `near-horizon' solution
with $a=0$! This result can be interpreted as a dominance of the M-Wave over
either  the M2-brane or the M5-brane in the
$r\rightarrow\infty$ limit because the static M2/M5 solution is not
asymptotically flat when $a=0$ whereas the M-wave is. Thus, the dominance
hierarchy for $r\rightarrow 0$ is precisely reversed when $r\rightarrow
\infty$.

\section{Discussion}

We have now discussed both the $r\rightarrow 0$ limit and the $r\rightarrow
\infty$ limits in both the generic case, and all special cases,
assuming that $a=0$, i.e. that $H=1/r^3$.
The various special cases that arise can be understood as particular
features of the generic solution in some characteristic range of the radial
coordinate $r$. Suppose that all charges $\xi_1,\xi_2,y,p$ are non-zero but
that one membrane charge is much larger than the other three
charges; this corresponds to the critical limit of constant $H$ on the
M5-brane. In this
case we expect the solution to look like that of the ($a=0$ and
$T^3$-compactified) M2-brane for $r$ not too small or large, i.e. the DW
solution. However, for sufficiently small $r$ the M5-brane will
dominate and the solution must approach the $adS_7\times S^4$ KK vacuum. On the
other hand, for sufficiently large $r$ the MW will dominate and the solution
must go to the flat D=11 vacuum. Thus, the dominance hierarchy translates to a
sequential interpolation from this flat D=11 vacuum at $r\approx \infty$ to the
DW spacetime at $r\sim 1$ and then on to the $adS_7\times S^4$ KK vacuum at
$r\approx 0$. This sequential interpolation corresponds to RG flow from
some 11-dimensional theory in the extreme UV (presumably M-theory) to the (2,0)
D=6 SCFT in extreme IR, passing through some intermediate theory which,
by the QFT/DW correspondence \cite{BST}, is presumably some D=3 field theory on
the D=4 Domain Wall.

For the special case of an M2-MW bound state with zero M5 charge, the dimension
of the transverse space jumps from 5 to 7. We can then choose $H$ to be a
harmonic function on this 7-space, and the simplest choice is
$H= 1/\rho^5$, where $\rho$ is the radial distance from the origin of $\bE^7$.
For the pure M2-brane this yields the $adS_4\times S^6 \times S^1$ DW solution
discussed in \cite{BST} as the near-horizon limit of the IIA D2-brane. This is
also the asymptotic spacetime as $r\rightarrow 0$ in
the `mixed' M2-MW case, but
in that case the $r\rightarrow \infty$ limit yields
a flat D=11 vacuum. This may
correspond to RG flow from the D=11 theory to a non-conformal D=3 field theory
on the D2-brane, but in this case one expects the extreme IR limit to be a
conformal D=3 field theory on the M2-brane, dual to the $adS_4\times S^7$
vacuum. This suggests that the choice $H=1/\rho^5$ of the harmonic function
is special, and that it could be replaced by a more general harmonic
function on
$\bE^7\times S^1$. Note that no analogous issue arises when the M5-brane
charge is non-zero because whereas the M-Wave direction is orthogonal to the
M2-brane it is parallel to the M5-brane.

\vskip 1cm
\noindent
{\bf Acknowledgements}: NO and PKT thank the University of Tokyo for
hospitality at the SI2000 Fujiyama workshop, where this work was initiated.
EB thanks DAMTP for hospitality.
The work of RGC and NO was supported in part by Grants-in-Aid for Scientific
Research Nos. 99020, 12640270 and Grant-in-Aid on the Priority Area:
Supersymmetry and Unified Theory of Elementary Particles.

%\newpage
\newcommand{\NP}[1]{Nucl.\ Phys.\ {\bf #1}}
\newcommand{\AP}[1]{Ann.\ Phys.\ {\bf #1}}
\newcommand{\PL}[1]{Phys.\ Lett.\ {\bf #1}}
\newcommand{\CQG}[1]{Class. Quant. Gravity {\bf #1}}
\newcommand{\CMP}[1]{Comm.\ Math.\ Phys.\ {\bf #1}}
\newcommand{\PR}[1]{Phys.\ Rev.\ {\bf #1}}
\newcommand{\PRL}[1]{Phys.\ Rev.\ Lett.\ {\bf #1}}
\newcommand{\PRE}[1]{Phys.\ Rep.\ {\bf #1}}
\newcommand{\PTP}[1]{Prog.\ Theor.\ Phys.\ {\bf #1}}
\newcommand{\PTPS}[1]{Prog.\ Theor.\ Phys.\ Suppl.\ {\bf #1}}
\newcommand{\MPL}[1]{Mod.\ Phys.\ Lett.\ {\bf #1}}
\newcommand{\IJMP}[1]{Int.\ Jour.\ Mod.\ Phys.\ {\bf #1}}
\newcommand{\JHEP}[1]{J.\ High\ Energy\ Phys.\ {\bf #1}}
\newcommand{\JP}[1]{Jour.\ Phys.\ {\bf #1}}

\end{document}